\documentclass[sigconf]{acmart}

\AtBeginDocument{%
  \providecommand\BibTeX{{%
    \normalfont B\kern-0.5em{\scshape i\kern-0.25em b}\kern-0.8em\TeX}}}

\begin{document}

\copyrightyear{2019} 
\acmYear{2019} 
\setcopyright{othergov}
\acmConference[CASA '19]{Computer Animation and Social Agents}{July 1--3, 2019}{PARIS, France}
\acmBooktitle{Computer Animation and Social Agents (CASA '19), July 1--3, 2019, PARIS, France}
\acmPrice{15.00}
\acmDOI{10.1145/3328756.3328758}
\acmISBN{978-1-4503-7159-9/19/07}

\title{Design of Seamless Multi-modal Interaction Framework for Intelligent Virtual Agents in Wearable Mixed Reality Environment }

\author{Ghazanfar Ali}
\affiliation{%
  \department{Division of NT-IT}
  \institution{University of Science and Technology}
   \country{South Korea}
}
\email{aliust@ust.ac.kr}

\author{Hong-Quan Le}
\affiliation{%
  \institution{Letsee}
   \country{South Korea}
}
\email{quan.le1926@gmail.com }

\author{Junho Kim}
\affiliation{%
  \department{College of Computer Science}
  \institution{Kookmin University}
   \country{South Korea}
}
\email{junho@kookmin.ac.kr}

\author{Seung-Won Hwang}

\affiliation{%
  \department{Department of Computer Science}
  \institution{Yonsei University}
   \country{South Korea}
}
\email{seungwonh@yonsei.ac.kr}

\author{Jae-In Hwang}
\authornote{Corresponding author}
\affiliation{%
  \department{Imaging Media Research Center}
  \institution{Korea Institute of Science and Technology}
   \country{South Korea}
}
\email{hji@kist.re.kr }

\renewcommand{\shortauthors}{G.Ali, et al.}

\begin{abstract}
   In this paper, we present the design of a multimodal interaction framework for intelligent virtual agents in wearable mixed reality environments, especially for interactive applications at museums, botanical gardens, and similar places. These places need engaging and no-repetitive digital content delivery to maximize user involvement. An intelligent virtual agent is a promising mode for both purposes.  Premises of framework is wearable mixed reality provided by MR devices supporting spatial mapping. We envisioned a seamless interaction framework by integrating potential features of spatial mapping, virtual character animations, speech recognition, gazing, domain-specific chatbot and object recognition to enhance virtual experiences and communication between users and virtual agents. By applying a modular approach and deploying computationally intensive modules on cloud-platform, we achieved a seamless virtual experience in a device with limited resources. Human-like gaze and speech interaction with a virtual agent made it more interactive. Automated mapping of body animations with the content of a speech made it more engaging. In our tests, the virtual agents responded within 2-4 seconds after the user query. The strength of the framework is flexibility and adaptability. It can be adapted to any wearable MR device supporting spatial mapping.
\end{abstract}


%\keywords{Mixed Reality, Virtual Characters, Interactive Virtual Agents,Multi-modal interaction framework}

\begin{teaserfigure}
\centering
  \includegraphics[width=0.7\textwidth]{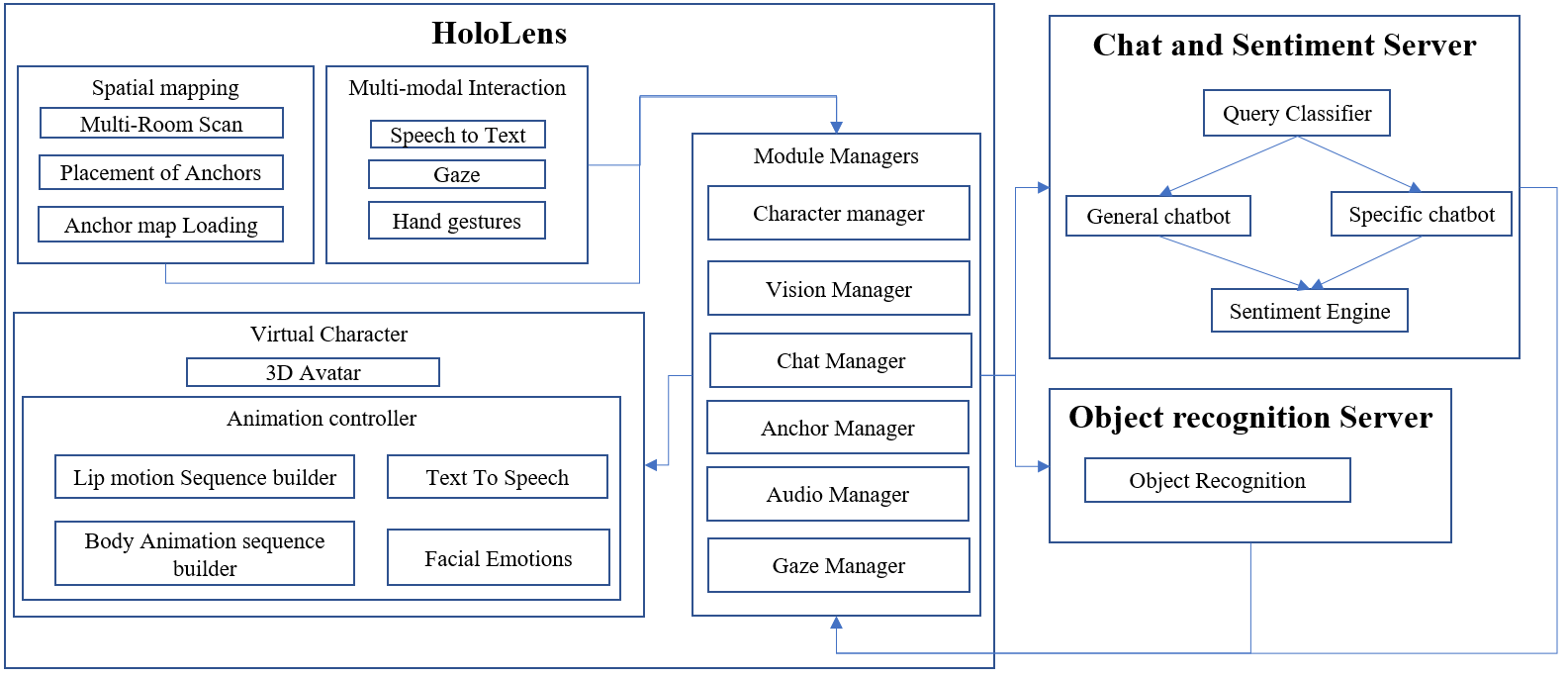}
  \caption{Overall framework for an intelligent virtual agent in wearable mixed reality.}
  \Description{Overview of the framework which explains the links between components. }
  \label{fig:teaser}
\end{teaserfigure}


\maketitle

\section{Introduction}
Mixed reality (MR) is a technique that blends the real and virtual worlds to create new environments where users can interact in real time. The ability to overlay the computer graphics images onto the real world with seamless interaction between virtual and physical worlds make MR technology suitable for wider ranges of the application as defined by Liu et al. \cite{RN1}. Presently, current-generation commercial mixed reality devices, such as Microsoft HoloLens, Magic Leap, and Meta 2, drive new methods of user interactions via hand gestures, voice, and gazing argues Xiao et al. \cite{RN2}. For decades, Agent-based systems have gained researcher's attention because of their potential for diverse applications as described by McArthur et al. \cite{RN3}, McArthur et al. \cite{RN4}, Catterson et al. \cite{RN5}, Castano et al. \cite{RN6} and Alencar et al. \cite{RN7}. In the paper of Xie et al. \cite{RN8}, significant features of agent-based systems were defined to emphasize its advantages. Anabuki et al. \cite{RN9} introduced a new type of anthropomorphic agent that lives in a 3D space where the real and virtual worlds are seamlessly merged. In research of Kopp et al. \cite{RN10}, a multimodal virtual humanoid agent assistant was created to have face-to-face discussions with users in a virtual environment. Ideally, the agent should mimic human behavior. Work of Cai et al. \cite{RN11} models human behavior for virtual agents. Machidon et al. \cite{RN12} provide a comprehensive review of the current state of the art in the use of virtual humans in cultural heritage ICT applications. Vosinakis et al. \cite{RN13} argue that virtual humans are more useful in the dissemination of intangible cultural heritage in virtual environments.  Today, many studies like  Arroyo-Palacios et al. \cite{RN14} and Bogdanovych et al. \cite{RN15} focus on making a believable character or agent partially show emotions and behavioral abilities when interacting with the environment or users. Unfortunately, this does not seem to be enough to enhance the interactive user experience. Richards \cite{RN16} presented a very detailed current study regarding agent-based architecture and intelligent virtual agent issues. However, most of these solutions are toward PC based systems and suffer from repetitive content delivery. The current generation of Wearable MR devices provides a more comfortable and engaging experience. It can further be enhanced by artistically crafted 3D virtual human, intelligent chatbot, object recognition, and intuitive interaction technique. Accordingly, our objective is the multimodal interaction of more than one agent in an MR environment, especially for museums and botanical gardens. We propose a framework which includes virtual character, multi-modal interaction, domain-specific chatbot, and object recognition system. The primary goal of this framework is to provide a seamless, interactive, engaging, and non-repetitive experience. The major strengths of our framework are its flexibility and adaptability as we utilized and modular approach. This framework gives the user the ability to experience the real world with digital information being delivered through a friendly virtual human. This gives the feeling of having a personal tour guide to enrich one's experience. Our contributions are:
\begin{itemize}
\item Design and implementation of the virtual agent framework for wearable MR.
\item Interaction design and implementation on the intelligent virtual agent
\item Expression and body gesture mapping of intelligent virtual agent
\item Scalable anchor system for wearable MR
\end{itemize}
The framework is explained in section 2. In section 3, we briefly introduce the development platform. In section 4, a scenario is presented and discussed.

\begin{figure}[ht]
  \centering
  \includegraphics[width=\linewidth]{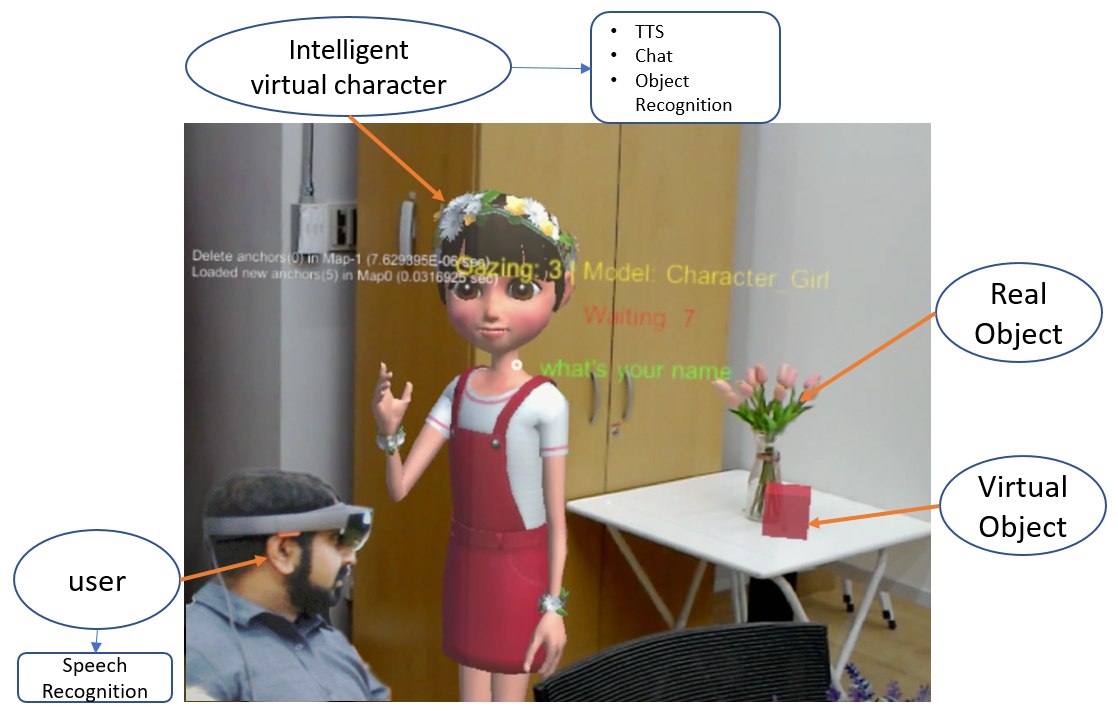}
  \caption{Scenario usage of the framework.}
  \Description{Usage scenario of framework with character and objects.}
  \label{fig:figure2}
\end{figure}

\section{Seamless MR System Architecture for Intelligent Virtual Agent}
\begin{figure}[ht]
  \centering
  \includegraphics[width=0.7\linewidth]{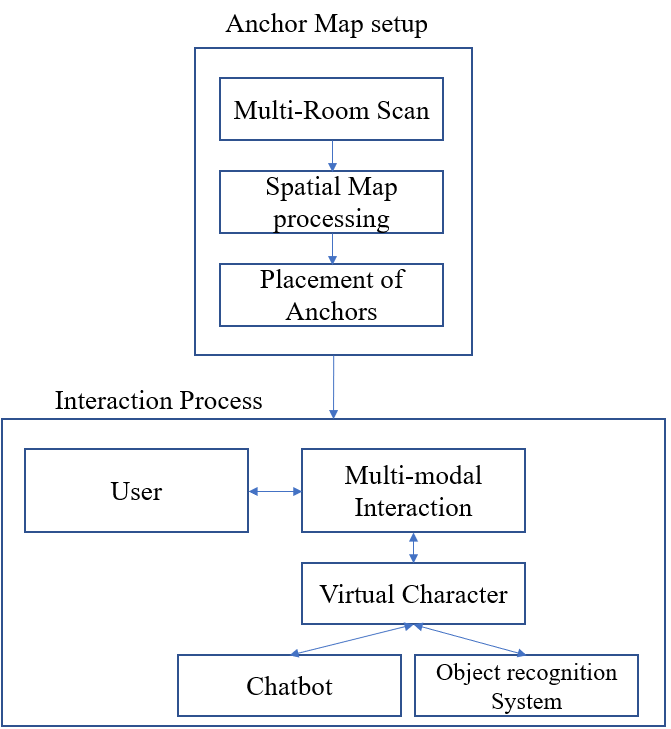}
  \caption{Overall architecture.}
  \Description{overview of the system architecture}
  \label{fig:figure3}
\end{figure}

To build a useful and meaningful system by integrating all required components, requirements in the wearable mixed reality system are:
\begin{itemize}
\item \textbf{Interaction:} Virtual agent which recognize user gazing objects and converse about it.
\item \textbf{Expression:} Virtual agent which shows emotion and body gesture which is appropriate to the conversation.
\item \textbf{Scalability:} Virtual agent who shows up at an infinite working area.
\end{itemize}
The overall architecture is shown in Figure \ref{fig:figure3}.

\subsection{Interaction: Gazing object recognition and conversation.}
\begin{figure}[ht]
  \centering
  \includegraphics[width=0.7\linewidth]{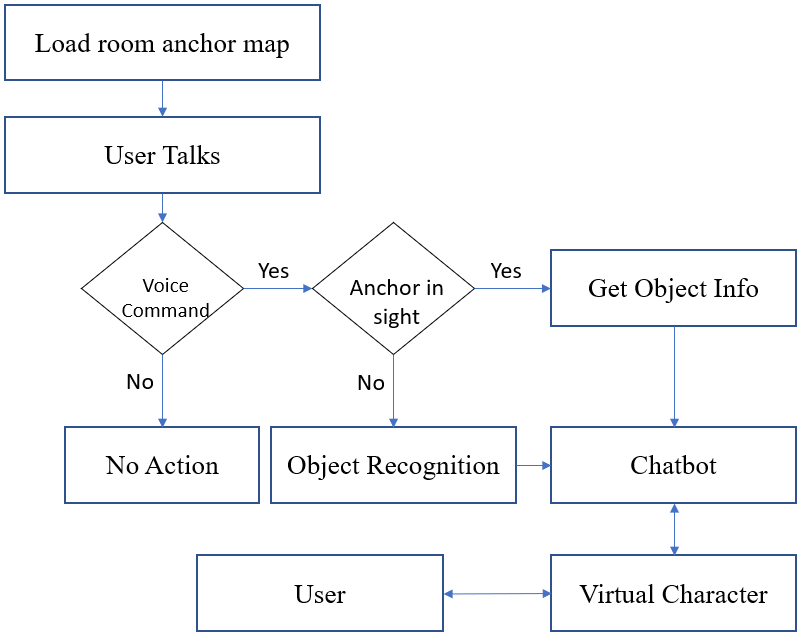}
  \caption{Process to get recognized object information}
  \Description{Usage of anchor map}
  \label{fig:figure4}
\end{figure}

The interaction process is aimed at providing a more natural method to interact with the virtual character. The major problem in the process was to figure out the way to ask the virtual character about any object for the first time. The problem arises when the user needs to ask about some object. If the user gazes at the character and asks about the object, the user must quickly look at the object to capture it. This sometimes does not give enough time to react. This problem was solved by including limited voice commands, i.e., "What is this," "Tell me about this" etc. in our framework. Figure \ref{fig:figure4} shows the interaction process with and without anchors in sight.

\subsubsection{Gaze and speech based multimodal interaction:}
There has been plenty of research about creating natural methods for human-agent interaction. For example, this study by Freigang et al. \cite{RN17} attempted to emphasize the role of speech and hand gestures in conveying pragmatic information between users and virtual humans. Arroyo-Palacios et al. \cite{RN14}  proposed an approach for interactive virtual characters for MR applications. Gaze and gestures were applied by users so that they could feel comfortable with a natural way to interact with agents. According to Kolkmeier et al. \cite{RN18}, gaze and proxemics behaviors can establish and maintain a level of intimacy, not only for human-human interaction but also for human-agent interaction. Results indicated that agents exhibiting higher proximity caused participants to step away from more than agents with low proximity.
\begin{figure}[ht]
  \centering
  \includegraphics[width=0.7\linewidth]{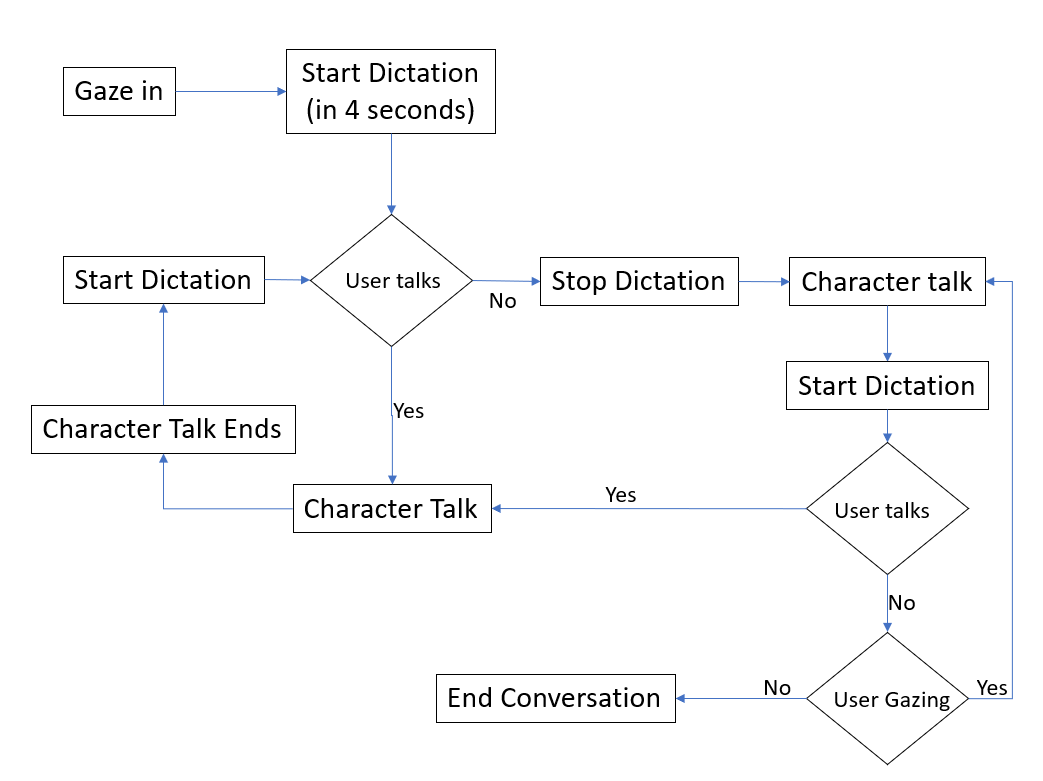}
  \caption{Gaze and Speech interaction architecture}
  \Description{Combination of gaze and speech for interaction}
  \label{fig:figure5}
\end{figure}
In a study by Yumak \cite{RN19}, a gaze behavior model was presented for an interactive virtual character that was situated in the real world. Uchino et al. \cite{RN20} and Rebman et al. \cite{RN21} showed the importance of speech recognition in interaction. The study of  Avramova et al. \cite{RN22} presented a toolkit with full-body animation for the virtual character, which offered MR interaction. We utilized gaze and speech recognition to interact with the virtual character. HoloLens provides the API for gaze and speech recognition. HoloLens Toolkit provides high-level functions of speech recognition like dictation manager, which enables us to say long sentences. Our method, as shown in Figure \ref{fig:figure5}, does not require any trigger keyword. When the user gaze on the virtual character for 4 seconds, speech recognizer starts. If the user talks, then it will wait to finish user talking and then send the query to a chatbot. If the user does not speak then, the virtual character will try to start the conversation by saying "Hello, do you need help?" or similar sentences. If the user gaze out while speech recognizer is running, and the user is silent, then the conversation will end.
\subsubsection{Gazed object recognition:}
Object recognition system works as the eyes of the virtual character. It enables the virtual character to be aware of its soundings. Any state of the art or custom trained object recognition or image classification system can be used if it can handle web requests. Vision manager module handles object recognition related task. It only needs a valid API Endpoint. The user can make their API Endpoint using any available system or use existing API Endpoint. Vision manager is flexible to handle as it can also be connected to the on-device system. By design, this can cover any number of objects if chatbot has the required knowledge base.

\subsubsection{Conversation of intelligent virtual agent:}
\begin{figure}[ht]
  \centering
  \includegraphics[width=0.7\linewidth]{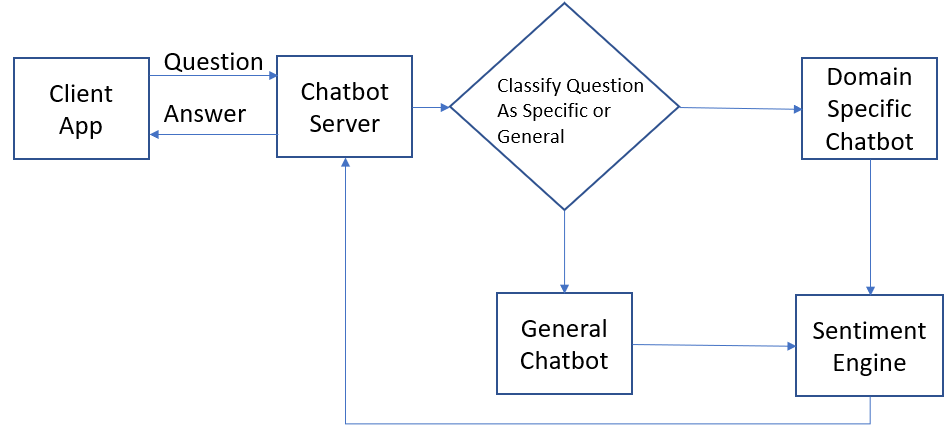}
  \caption{Chatbot Architecture}
  \Description{Chatbot overview}
  \label{fig:figure6}
\end{figure}

For a virtual character to interact with the user, it needs a knowledge base and associated program to give out the required information. Those programs are called chatbots. Chatbots are playing a vital role in Human-Computer Interactions. Chatbots are also being used to automate customer services. In our system, chatbot serves two purposes, i.e., 1) It provides domain-specific information. 2) General conversation options. Our chatbot is equipped with a sentiment engine which analyses the response and attaches a sentiment class and level of sentiment to it as shown in Figure \ref{fig:figure6}. This chatbot is deployed on the cloud and accessed through web requests. Input parameter to this chatbot are "(\textit{Query, Object})." \textit{Query} can be a question or comment or an answer to virtual character's question. \textit{Object} is any virtual or real object detected through object recognition about which we want to ask a question. \textit{Query} can be a general query like asking about the character itself or can be a specific query about the objects around you. Prerequisite for a specific query about any object is that the user must look at the object for the first time and ask about it. This will trigger the object recognition component and will pass the information to a chatbot. Follow up queries don't require to look at the object. Reply to the \textit{Query} is "(\textit{Reply, Sentiment Class, Sentiment Level})." Sentiment class and the level are used for appropriately animating the character. The user can ask both general or specific query in one conversation and any order. This style of conversation mimics human-human interaction. This increases engagement and improves the quality of experience. 

\subsection{Expression: Virtual character emotion and body gesture}
\begin{figure}[ht]
  \centering
  \includegraphics[width=0.7\linewidth]{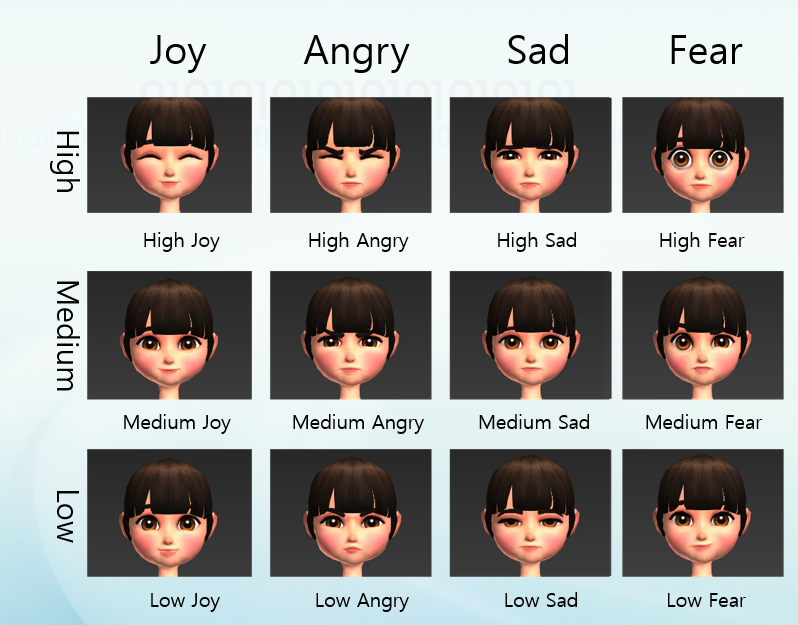}
  \caption{Facial emotions types and levels}
  \Description{Facial emotions and levels}
  \label{fig:figure7}
\end{figure}

Our virtual human is a 3D model. To make a realistic virtual human, it needs to have rich body animations and facial emotions. Body animations are for verbal and nonverbal activities. As virtual human is required to move around and respond to verbal cues, it needs to have a large set of animations available. In our system, we have 30 different animations. Facial emotions are for the expression of different emotions. In our system, we have four emotions, i.e., Joy, Angry, Sad, Fear. Each emotion has three levels, i.e., High, Medium, Low, as shown in Figure \ref{fig:figure7}.

\begin{figure}[ht]
  \centering
  \includegraphics[width=0.7\linewidth]{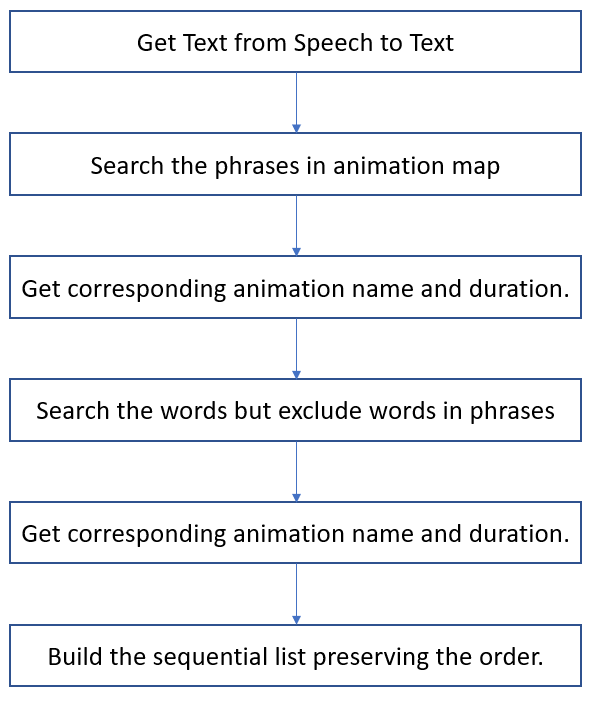}
  \caption{Architecture for building body-animation sequence}
  \Description{Process of building body animation sequence.}
  \label{fig:figure8}
\end{figure}
\begin{figure}[ht]
  \centering
  \includegraphics[width=0.7\linewidth]{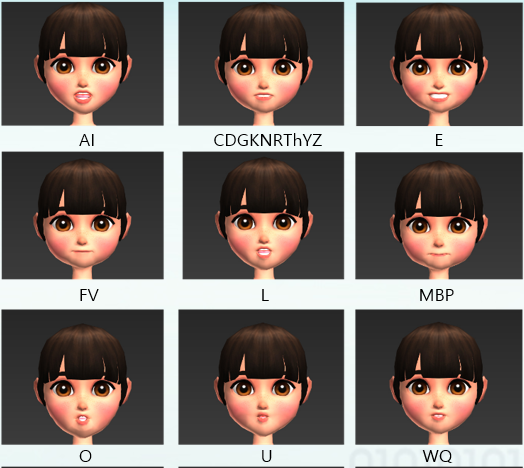}
  \caption{Lip shape for phonemes}
  \Description{all Lip shape for phonemes }
  \label{fig:figure9}
\end{figure}
\begin{figure}[ht]
  \centering
  \includegraphics[width=0.7\linewidth]{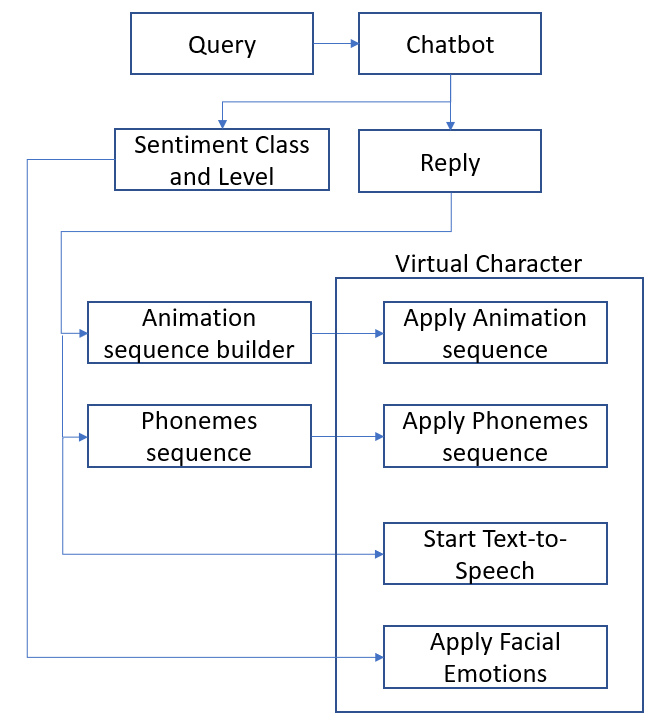}
  \caption{Applying speech, animation and facial emotions on the character}
  \Description{process of applying speech, animation and facial emotions.}
  \label{fig:figure10}
\end{figure}

Body animations and facial emotions are applied based on the content of the speech. Speech is generated from text using Text-to-Speech system. The architecture for building animation sequence is shown in Figure \ref{fig:figure8}. This architecture requires a mapping table for mapping of text to animations. Human experts create this map. It has an animation for phrases and individual words which are utilized by the animation builder, as shown in Figure \ref{fig:figure8}. Facial emotions are predicted from the text by sentiment engine integrated into chatbot and applied for the complete duration of the text. Speech from the text is lip-synced by converting text to phonemes sequence. Phonemes map is used for this purpose. Lip shape for each phoneme is shown in Figure \ref{fig:figure9}. Speech, body animation, facial emotions are applied in parallel, as shown in Figure \ref{fig:figure10}.

\subsection{Scalability: Anchor map setup}             
\begin{figure}[ht]
  \centering
  \includegraphics[width=\linewidth]{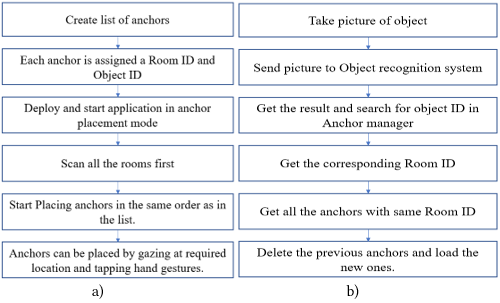}
  \caption{a) Anchor placement process. b) Anchor loading}
  \Description{process of anchor placement and loading process.}
  \label{fig:figure11}
\end{figure}

HoloLens provides a mechanism known as anchors through which static objects can be tracked. By combining image recognition and anchors, we created a process by which we can map infinite working area. This process can save information about real objects which can be accessed by the virtual agent. This process minimizes the use of object recognition for each object. The anchor placement process is shown in Figure \ref{fig:figure11} a). Initial two steps are part of the framework's anchor manager. Since there is no mechanism to load room-specific anchors in HoloLens and for identical rooms wrong anchors may be loaded. To avoid this issue, anchor management process utilizes object recognition to load room-specific anchors. The process is shown in Figure \ref{fig:figure11} b).

\section{Development Environment}
The platform is consisting of Unity3D, Microsoft HoloLens, HoloLens Toolkit, and cloud computing platform. HoloLens is spatial MR holographic device developed by Microsoft. With HoloLens, we scan the room to create a 3D space through which user, virtual objects, and virtual character are tracked. HoloLens supports Unity3D as a development platform. HoloLens toolkit provides an easy way of setting up initial project and settings. It also offers numerous helping functions with regards to mapping, gaze tracking, and speech recognition. HoloLens is a self-contained device with limited resources. It is not possible to efficiently run chatbot and object recognition from the device without compromising the quality of user experience. To solve this, our framework access cloud-based chatbot and object recognition system via module manager.

\section{Scenario: A Botanical garden}
The framework was utilized to create a demo system. The scenario was a tour at a Botanical garden. In this scenario, the user can interact with the virtual character to ask about different flowers. This scenario needed the following items:
\begin{itemize}
\item 3D humanoid model
\item Flower dataset for the object recognition system
\item Knowledgebase of flowers for chatbot
\end{itemize}
3D artists created 3D Models. It included body design, clothes, and facial emotions. Body animations were acquired from Mixamo[https://www.mixamo.com]. There were three models, i.e., Girl, Boy, and Old Man. Different voices from Text-To-Speech systems were assigned to them. Body animation also varies for each character. No customization was required in the framework to add multiple characters in one application. Categories of the flower were nine. We used artificial flower models. A dataset of images was prepared from those models. Our object recognition system was trained and deployed on custom vision API of Microsoft Azure. We trained the system using our dataset. Knowledgebase for chatbot was built with extensive research. The goal for chatbot was to include as much information as possible so that the user should not be forced to ask only specific questions.
\begin{figure}[ht]
  \centering
  \includegraphics[width=\linewidth]{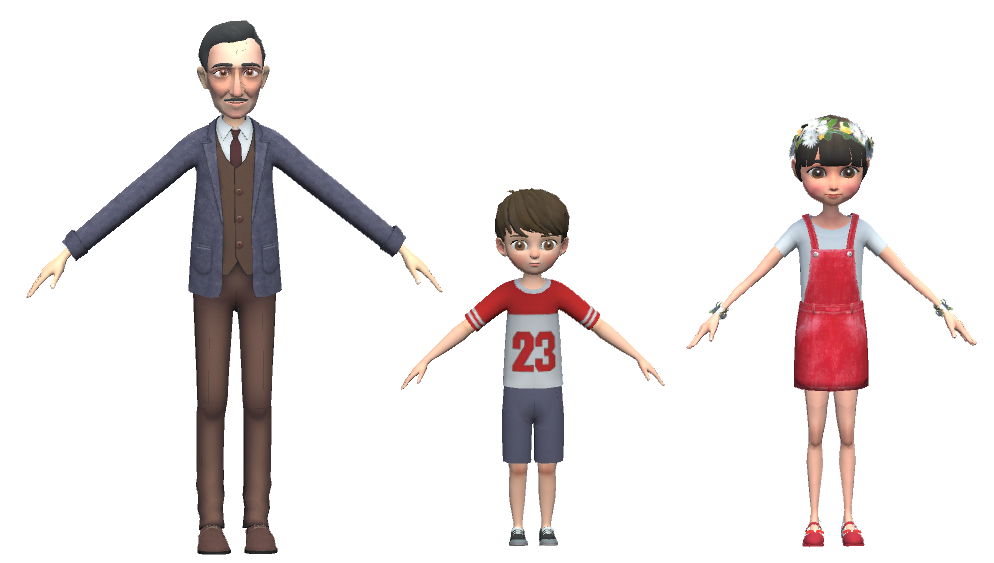}
  \caption{Models}
  \Description{3D Models used in scenario i.e. Old man , boy , girl}
\end{figure}
\begin{figure}[ht]
  \centering
  \includegraphics[width=\linewidth]{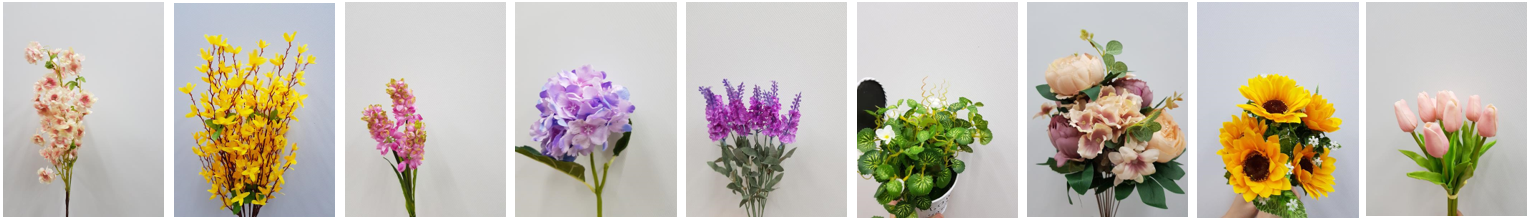}
  \caption{Nine flowers used in the scenario.}
  \Description{Flower data set}
\end{figure}
\begin{figure}[ht]
  \centering
  \includegraphics[width=\linewidth]{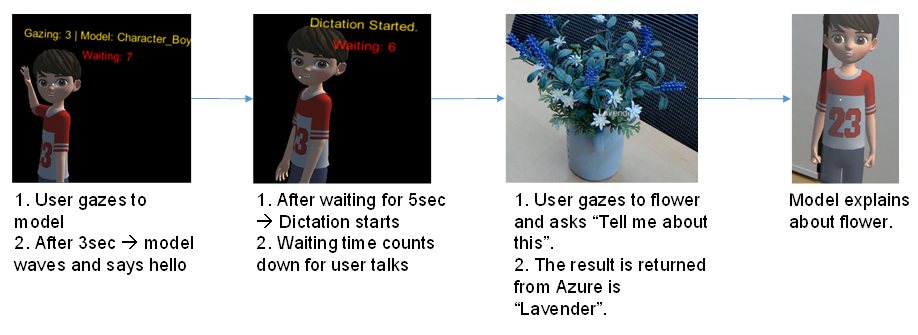}
  \caption{Interaction with model}
  \Description{demo of interaction with model for specific conversation}
\end{figure}

For this scenario, two rooms were set up with five flowers in 1 room and 4 in 1 room. Scanning was performed, and anchors were placed. Girl model was assigned room 1, and Boy model was assigned room 2. Then we successfully demonstrated the framework. The outcome was a seamless interaction with response time ranging between 2-4 seconds. This time includes speech-to-text conversion, sending text query to a chatbot, receiving a response, and building animation list. This response time, however, does not include the time of silence to detect the end of user speech. The response time will be higher if object recognition is also triggered typically in the range of 5-8 seconds total. As this delay is more significant, so during this time, our character shows the animation of thinking about the object and say something like "let me see, let me think about it." This way, the user does not feel a time delay. As our modules were deployed on cloud-platform and access through the internet, so a significant portion of the time was utilized in web requests. In Table \ref{tab:timing}, we summarize the average results for three queries. Query A was loading of anchor map, and it required object recognition. Query B was general conversation. Query C was about flower, and it needed object recognition.
\begin{table}
  \caption{Total average response time for each query}
  \label{tab:timing}
  \begin{tabular}{|c|c|c|c|}
  
    \toprule
    Query&Query A&Query B&Query C\\
    \midrule
    Total queries&30&30&30\\

   Object recognition&5.4s&n/a&5.3s\\
  
   Chatbot&n/a&2.1s&2.0s\\
 
   Processing time&0.5s&1s&1s\\

   Total Time&5.9s&3.1s&8.3s\\

   Std Dev&0.99&0.98&0.99\\
  \bottomrule
\end{tabular}
\end{table}
Above table proves the usability of anchor map to reduce response time.
\section{CONCLUSIONS}
We studied and conceptualized a framework by integrating spatial mapping, virtual character, multi-modal interaction, chatbot, and object recognition for wearable mixed reality. We successfully demonstrated that such a framework could reduce the time to develop an augmented reality environment with an intelligent virtual agent. It can prove to be more engaging and entertaining. The approach presented both theoretical and experimental aspects to emphasize the importance of using multimodal interactions for humans and agents. The potential of spatial mapping, virtual character, multi-modal interaction, chatbot, and object recognition was effectively integrated to enhance the human-agent interaction and to stimulate the realism of the mixed reality environment.

\begin{acks}
This research is supported by the Korea Creative Content Agency (KOCCA) in the Culture Technology (CT) Research and Development Program and KIST Flagship Project.
\end{acks}
%%%%%%%%%%%%%%%%%%%%%%%%%%%%%%%%%%%%%%%%%%%%%%%%%%%%%%%%%%%%%%%%%%%%%%%%%%%%%%%%%%%%%%%%%%%%%%%%%%%%%%%%%%%%

\bibliographystyle{ACM-Reference-Format}
\bibliography{MR-Framework}

\end{document}